\documentclass[pra,twocolumn,superscriptaddress,showpacs,10pt]{revtex4}

\usepackage{graphicx}
\usepackage{dcolumn}
\usepackage{bm}

\begin{document}


\title{Effective size of a trapped atomic Bose gas}

\author{Wenxian Zhang}
\affiliation{School of Physics, Georgia Institute of Technology,
Atlanta, Georgia 30332, USA}

\author{Z. Xu}
\affiliation{Center for Advanced Study, Tsinghua University,
Beijing 100084, People's Republic of China}

\author{L. You}
\affiliation{School of Physics, Georgia Institute of Technology,
Atlanta, Georgia 30332, USA}
\affiliation{Center for Advanced Study, Tsinghua University,
Beijing 100084, People's Republic of China}

\date{\today}

\begin{abstract}
We investigate the temperature-dependent effective size of a trapped
interacting atomic Bose gas within a mean field theory approximation.
The sudden shrinking of the average length, as observed
in an earlier experiment by Wang {\it et al.} [Chin. Phys.
Lett. {\bf 20}, 799 (2003)], is shown to be a good indication
for Bose-Einstein condensation (BEC).
Our study also supports the use of the average width of a trapped Bose gas
for a nondestructive calibration of its temperature.
\end{abstract}


\maketitle

\section{Introduction}

It has been a decade since the first realization of Bose-Einstein condensation
(BEC) in alkali atomic gases \cite{Anderson95, Davis95, Bradley95}. Many
properties of atomic quantum gases have been extensively studied and understood
\cite{WebAmo, PethickBook02, PitaevskiiBook03, Dalfovo99}. A consistent theme
has been the general agreement between theoretical predictions and experimental
observations. For example, at low temperatures when the non-condensed thermal
component is negligible, Gross-Pitaevskii equation has been successfully used
to model the condensate mean field, which provides a firm theoretical framework
for the description of condensate density distribution and collective
excitations \cite{Dalfovo99,Salasnich00, Edwards95, Ruprecht95, Edwards96,
Krauth96, Dalfovo96, Mewes96, Holland96, Holland97, Matthews98}. At nonzero
temperatures, several mean field theory extensions such as the
Hartree-Fock-Bogoliubov or the Hartree-Fock-Popov approaches generally become
sufficient, as evidenced by the calculated temperature-dependent condensate
fraction, expansion energy, and number fluctuation, etc. \cite{Goldman81,
Giorgini96, Giorgini97, Shi97, Bagnato87, Dalfovo97, Tanatar02, Grossmann97,
Anderson99,Holzmann99, Idziaszek99, Xiong02, Liu04, Lewenstein94, Ketterle96,
Ensher96, Gerbier04, Chou96}.

In this paper we report our investigations of the temperature-dependent
average width of a trapped interacting Bose gas within the mean field
Hartree-Fock approximation. We undertake this study in an
effort to provide some theoretical support for the recent experiment by Wang
{\it et al.}, which produced the first atomic condensate
inside mainland China \cite{Wang03}.
Few earlier theoretical studies have addressed the
issue of an averaged length/width of a trapped interacting Bose gas
\cite{Giorgini97}, despite it being one of the topics studied in the first
generation of BEC experiments \cite{Andrews96}. As expected, our study
points to an universal feature for all trapped interacting Bose gases.
We find that the effective, or the
average size of a repulsively interacting trapped Bose gas drops suddenly when
condensation occurs, i.e. after the gas temperature is lower than the
transition temperature $T_C$. As a consequence we support the use of this
reduction in effective sample size as a good evidence for the onset of BEC,
distinctively different from the more conventional bimodal density
profile based on
a measurement of the expanded condensate density distribution
after the confining trap is switched off.

This paper is organized as follows. The present section provides a brief
introduction. Sections \ref{sec:si} and \ref{sec:st} describe the calculations
for the average size of an ideal or interacting trapped Bose gas respectively.
We analyze the energy partitions of a Bose gas in Sec. \ref{sec:en}. Section
\ref{sec:cl} presents a short conclusion which summarizes our results.

\section{Size of an ideal Bose gas}
\label{sec:si}

For an ideal Bose gas, the average number of particles in a
single particle state $|i\rangle$ with energy
$\varepsilon_i$ is given by the familiar Bose-Einstein distribution
\begin{eqnarray}
n_i &=& \frac {g_i} {e^{\beta(\varepsilon_i-\mu)}-1} =  g_i \;\frac {{\sf z}
e^{-\beta \varepsilon_i }} {1-{\sf z}e^{-\beta\varepsilon_i}},
\end{eqnarray}
where $\beta=1/(k_BT)$ with $k_B$ denoting the Boltzmann constant, $g_i$ is the
degree of degeneracy for state $|i\rangle$, ${\sf z}=\exp(\beta\mu)$ is the
fugacity, and $\mu$ is the chemical potential determined by
the conservation of total number ($N$) of particles
$\sum_{i=0}^\infty n_i = N.$

The statistical properties of a Bose gas, such as its specific heat, the
condensate fraction, etc, are completely determined once the chemical potential
is found \cite{Bagnato87, Lewenstein94, Ketterle96, Giorgini96, Gerbier04}. For
a spherically symmetric harmonic trap $V_{\rm ext}(r) = M\omega^2 r^2/2$, the
width for state $|i\rangle$ is easily found to be
\begin{eqnarray}
\langle r_i^2 \rangle &=& \frac {\varepsilon_i} {M \omega^2} = \left(i+{3\over
2}\right) a_r^2,
\end{eqnarray}
where $a_r =\sqrt{\hbar/(M\omega)}$ is a characteristic length for the harmonic trap.
The width of a Bose gas is then obtained as
\begin{eqnarray}
\langle r^2\rangle &=& \left(\frac 32 a_r^2\right)\frac {\sf z} {1-{\sf z}} +
\sum_{i=1}^\infty n_i\left(i+ \frac 32 \right) a_r^2.
\end{eqnarray}
For large N, we take the usual approximation of changing the summation
into an integral weighted by the density of states and
separate out the atoms in the ground state. We find that
\begin{eqnarray}
\langle r^2\rangle &\simeq & \left(\frac 32 a_r^2\right)N_0 + 3a_r^2\; \left(
\frac {k_BT}{\hbar\omega}\right)^4\; g_4({\sf z}), \label{eq:sw} 
\end{eqnarray}
if the temperature satisfies $k_BT\gg\hbar\omega$, where
\begin{eqnarray}
N_0 &=& \left\{\begin{array}{ll}
0, & T\ge T_C, \\
N-\left(k_BT\over \hbar\omega\right)^3 g_3({\sf z}), & T<T_C, \end{array}
\right. \nonumber
\end{eqnarray}
with $g_\nu({\sf z}) = \sum_{k=1}^\infty {\sf z}^k/k^\nu$,
$\zeta(\nu)=g_\nu(1)$, and $T_C=(\hbar\omega/k_B)(N/\zeta(3))^{1/3}$. This
expression (\ref{eq:sw}) takes a familiar form with the first term denoting the
squared width for the ground state (condensate), while the second term for the
excited states (thermal component).

\begin{figure}
\begin{center}
\includegraphics[width=3.35in]{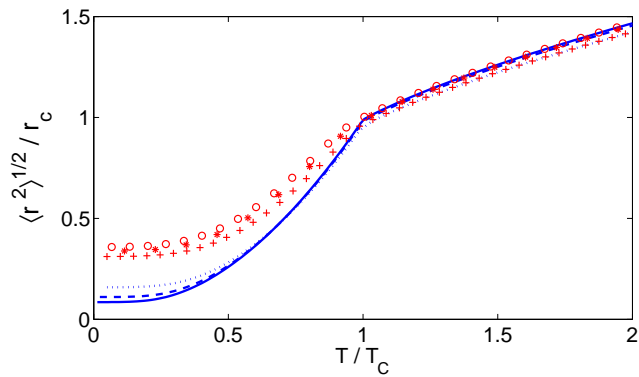}
\caption{(Color online) The temperature-dependent width of a Bose gas in a
spherically symmetric harmonic trap. The solid, dashed, dotted lines denote
respectively the case of an ideal gas with $N=5\times 10^5$, $1\times 10^5$,
and $1\times 10^4$. The corresponding results for an interacting gas are given
by empty circles, asterisks, and plus signs. The trap parameters are given in
Table \ref{table:ttp}. Note that $T_C$ and $r_C$ both depend on $N$.}
\label{fig:fig1}
\end{center}
\end{figure}

We note there exist two apparent temperature scales in the system,
$\hbar\omega$ and $k_BT_C$. When $k_BT_C\gg \hbar\omega$, as normally is the
case, our approximation is expected to work well. For high temperatures
$k_BT\gg\hbar\omega$, we ignore the contribution from the condensate and obtain
\begin{eqnarray}
\langle r^2\rangle &\approx & 3a_r^2\; \left(\frac {k_BT}
{\hbar\omega}\right)^4\; \zeta(4) = r_C^2 \left({T\over T_C}\right)^4,
\end{eqnarray}
for $T<T_C$. On the other hand, we find
\begin{eqnarray}
\langle r^2\rangle &\approx & 3a_r^2N\left(\frac{k_BT}{\hbar\omega}\right)
{g_4({\sf z})\over g_3({\sf z})} = \alpha\, r_C^2 \left({T\over T_C}\right),
\end{eqnarray}
for $T>T_C$. $r_C^2=3a_r^2\zeta(4) [N/\zeta(3)]^{4/3}$ denotes the squared
width at $T_C$ and $\alpha = g_4({\sf z}) \zeta(3) / [g_3({\sf z}) \zeta(4)]
\sim 1$ has a very weak dependence on $T$. We see that in
a spherically symmetric harmonic trap, the squared width of
an ideal Bose gas is proportional to
$T^4$ for $T<T_C$ and to $T$ for $T> T_C$, as clearly
shown in Fig. \ref{fig:fig1}, consistent with earlier experimental reports
that the area of absorption image of a Bose gas is proportional to its
temperature in the absence of a condensate \cite{Andrews96, Wang03}.

\begin{figure}
\begin{center}
\includegraphics[width=3.35in]{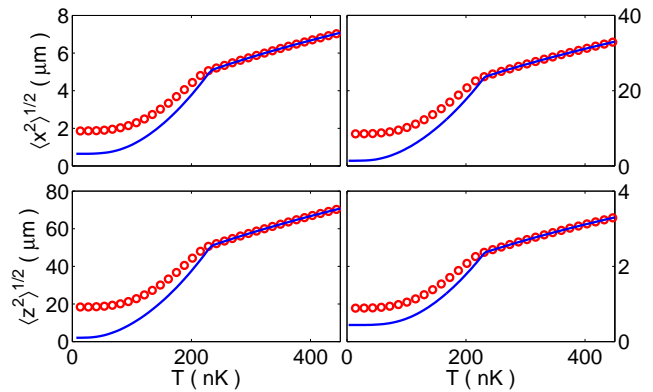}
\caption{(Color online) The same as in Fig. \ref{fig:fig1}
but for a cigar-shaped harmonic trap (the left column)
or a pancake-shaped harmonic trap (the
right column). The total number of atoms is assumed $N=5\times 10^5$ with trap
parameters as given in Table \ref{table:ttp}. The solid lines are for
an ideal gas while the empty circles refer to an interacting
$^{87}$Rb gas.} \label{fig:fig2}
\end{center}
\end{figure}

For a cylindrically symmetric trap with $\omega_x=\omega_y=\omega_\perp$,
the temperature dependence of the three widths squared is the same
as in a spherically symmetric trap discussed above. The prefactors
become respectively $z_C^2 = a_z^2 \lambda^{-2/3}
\zeta(4)[N/\zeta(3)]^{4/3}$ and
$x_C^2 = y_C^2 = a_\perp^2 \lambda^{1/3}
\zeta(4)[N/\zeta(3)]^{4/3}$ for the axial and transverse directions (Fig. \ref{fig:fig2}),
 assuming an axial trap frequency $\omega_z=\lambda \omega_\perp$.
$a_z=\sqrt{\hbar/(M\omega_z)}$ and $a_\perp=\sqrt{\hbar/(M\omega_\perp)}$ are
the respective characteristic lengths for the harmonic trap along the axial and
transversal directions. If we approximate the effective area from the side view
of a Bose gas as $S\approx \sqrt{\langle z^2\rangle \langle x^2\rangle}$, we
find the temperature dependence of $S$ satisfies
\begin{eqnarray}
S=\left\{\begin{array}{ll}
S_C \left({T\over T_C}\right)^4, & T \le T_C, \\
\alpha S_C \left({T\over T_C}\right), & T \ge T_C, \end{array} \right.,
\end{eqnarray}
with $S_C = \sqrt{z_C^2 x_C^2}$\,. Once again, this is clearly consistent
with earlier observations that the effective area is proportional to
temperature above BEC and drops suddenly below condensation temperature
\cite{Andrews96, Wang03}.

\section{Size of an interacting Bose gas}
\label{sec:st}

With atomic interactions, the Hamiltonian of our model system becomes
\begin{eqnarray}
H &=& \int d\vec r \left[ \Psi^\dag\left(-{\hbar^2\over 2M}\nabla^2+V_{\rm
ext}(\vec r)\right) \Psi+{g\over 2}\Psi^\dag\Psi^\dag\Psi\Psi\right], \nonumber
\end{eqnarray}
in the second quantized form,
where $\Psi(\vec r)[\Psi^\dag(\vec r)]$ is the quantum field for annihilating
(creating) an atom at location $\vec r$
and $g=4\pi\hbar^2a_s / M$ denotes the interaction strength
with $a_s$ being the $s$-wave scattering length. It is easy to show that the
total number of atoms $N$ is conserved since it commutes with the Hamiltonian.
We then introduce a Lagrange multiplier $\mu$ to fix the total number of atoms
in our numerical calculations. In fact, $\mu$ is simply the chemical potential.
The target function we need to minimize then becomes the free energy
\begin{eqnarray}
G &=& H-\mu N \nonumber \\
&=& \int d\vec r \left[ \Psi^\dag\left(-{\hbar^2\over 2M}\nabla^2+V_{\rm
ext}(\vec r)-\mu\right) \Psi\right. \nonumber \\
&& \left. + {g\over 2}\Psi^\dag\Psi^\dag\Psi\Psi\right].
\end{eqnarray}
Adopting the standard mean field theory \cite{Zhang04, Giorgini97},
we introduce a condensed ($\Phi=\langle \Psi\rangle)$
and a thermal component of a Bose gas through
$\Psi = \Phi + \delta \Psi$. After straightforward manipulations, we
obtain the Gross-Pitaevskii equation for the two components as,
\begin{eqnarray}
i\hbar \frac {\partial}{\partial t}\Phi(\vec r, t) &=& \left[-{\hbar^2\over
2M}\nabla^2 + V_{\rm ext} - \mu + g(n+n_T)\right]\Phi, \nonumber \\
\label{eq:gpe}\\
i\hbar \frac {\partial}{\partial t}\delta\Psi(\vec r, t) &=&
\left[-{\hbar^2\over 2M}\nabla^2 + V_{\rm ext} - \mu+2gn\right]\delta\Psi,
\label{eq:te}
\end{eqnarray}
where $n=n_C+n_T$ is the total density with $n_C=|\Phi|^2$ being the condensate
density and $n_T$ the density of thermal atoms [given by Eq. (\ref{eq:t})]. In
obtaining the above closed set of equations, we have used the Hartree-Fock (HF)
approximation, which is both efficient for numerical solutions and reasonably
accurate for predicting the statistical properties of a Bose gas as confirmed
by the excellent agreement between experiments and theoretical results
\cite{Goldman81, Ensher96, Giorgini97, Shi97, Holzmann99, Xiong02, Gerbier04}.

We will treat the thermal component through a semi-classical approximation,
$-i\hbar \nabla \rightarrow \vec p$, which gives its distribution in terms
of a Bose-Einstein distribution in phase space \{$\vec p, \vec r$\}
\begin{eqnarray}
n_T(\vec r)&=&\int {d\vec p\over (2\pi\hbar)^3}{1\over e^{\varepsilon(\vec
p,\vec r)/k_BT}-1}, \label{eq:t}
\end{eqnarray}
with $\varepsilon(\vec p,\vec r)=p^2/(2M) + V_{\rm ext} -\mu+2gn$. Numerically
we follow the standard procedure solving for the self-consistent solution of
the coupled Eqs. (\ref{eq:gpe}), (\ref{eq:te}),
and (\ref{eq:t}) \cite{Zhang04, Giorgini97}.

\begin{table}
\caption{Parameters for various types of typical harmonic traps,
in units of ($2\pi$)Hz. The cigar-shaped trap is for the experiment of Ref. \cite{Wang03}.}
\begin{tabular}{l|rrr}
\hline
\hline
&$\omega_x$&$\omega_y$&$\omega_z$\\
\hline
Spherically symmetric    &65&65&65\\
Cigar-shaped &140&140&14\\
Pancake-shaped &30&30&300\\
\hline
\hline
\end{tabular}
\label{table:ttp} 
\end{table}

We compute below the temperature dependent mean field ground state of an
interacting $^{87}$Rb atomic Bose gas inside a harmonic trap
\begin{eqnarray}
V_{\rm ext}(x,y,z) &=& {1\over 2} M (\omega_x^2x^2 + \omega_y^2y^2 +
\omega_z^2z^2 ),
\end{eqnarray}
taking the atomic $s$-wave scattering length $a_s = 100.4 a_B$ with $a_B$ the
Bohr radius \cite{para2}. For spherically symmetric traps, we search for the
ground state with different total number of atoms, $N=5\times 10^5$, $1\times
10^5$, and $1\times 10^4$. The temperature dependence of the effective width,
defined as $\langle r^2\rangle = \int d\vec r \;r^2 n(\vec r)$, is shown in
Fig. \ref{fig:fig1}. Clearly a sudden drop of the effective width for an atomic
cloud occurs when temperature is lower than the critical temperature, as in the
experiments \cite{Andrews96, Wang03}. We further observe that the sudden
decreasing of the reduced width near the critical temperature has little
dependence on the number of atoms.

Figure \ref{fig:fig1} also reveals that the difference between an ideal gas and
an interacting one increases with the number of atoms, and with increasing
repulsive interaction strength for $T<T_C$. An interesting feature we note is
that the repulsive interaction causes the width of a Bose gas to increase at
temperatures lower than the critical temperature ($T<T_C$), yet it has little
effect on the width at temperatures higher than the critical temperature
($T>T_C$). The low temperature phenomenon is easy to understand in terms of a
repulsive-interaction induced expansion of a Bose gas \cite{Giorgini97}. First,
a condensate with repulsive interaction is larger in its size due to atom-atom
repulsion; Second, the presence of a condensate pushes the thermal
non-condensed cloud out, further increasing the width of a gas \cite{Liu04,
Naraschewski98}. At high temperatures ($T>T_C$) the effect of repulsive
interaction becomes negligible as the density of a Bose gas decreases
dramatically with increasing temperatures.

For cylindrically symmetric harmonic traps, we computed
the ground states both for a cigar-shaped harmonic trap
with an aspect ratio $\omega_z / \omega_x = 0.1$ and a pancake-shaped
harmonic trap with
$\omega_z/\omega_x=10$. As before, we used $N=5\times 10^5$.
Of particular interest, we find excellent agreement with the
 experimental observations of Wang {\it et al.}
\cite{Wang03} for the above cigar shaped trap, at approximately the same
transition temperature $T_C\approx 230 n$K for their experimental parameters.
In the experiment of Wang {\it et al.} \cite{Wang03}, the magnetic trap cannot
be switched off without causing violent perturbations to trapped atoms.
Therefore, {\it in situ} near resonance imaging \cite{You} was used instead of
the conventional approach of shutting off the trap and letting a condensate
expand \cite{Holland96}. After condensation, a characteristic halo-like
structures was observed in near resonantly diffracted light \cite{Bradley95}.
Normally this cannot be taken as a convincing evidence for BEC. But for a cigar
shaped trap as was used in their experiment, the halo is mainly along the
transverse direction, while the axial length can still be determined with
sufficient accuracy because it is relatively long, in fact much longer than the
resonant optical wavelength; When the temperature dependence of the effective
length for their experiment was plotted, a sudden decrease was observed, which
was interpreted as an indication of the BEC phase transition \cite{Wang03}. Our
work thus provides a solid theoretical foundation for their experiment. As
illustrated in the above figures, a sudden decrease of the effective length of
the cloud was observed when the temperature is lower than the transition
temperature, thus a corresponding decrease of the effective area. Similar to
spherically symmetric traps, the difference between the effective widths of an
interacting Bose gas and an ideal gas decreases when the temperature increases
and becomes negligible for $T>T_C$. We also note that the ratio between the
transverse and the axial width remains about the same as the aspect ratio
$\omega_z / \omega_x$ for the interacting Bose gases. This is not surprising
because this ratio approaches that of an ideal gas $x_C/z_C=\omega_z/\omega_x$
when $T>T_C$ or that of the TF limit $x_{TF}/z_{TF}=\omega_z/\omega_x$ when
$T\to 0$.

\section{Partition of Energy for an interacting Bose gas}
\label{sec:en}

\begin{figure}
\begin{center}
\includegraphics[width=3.35in]{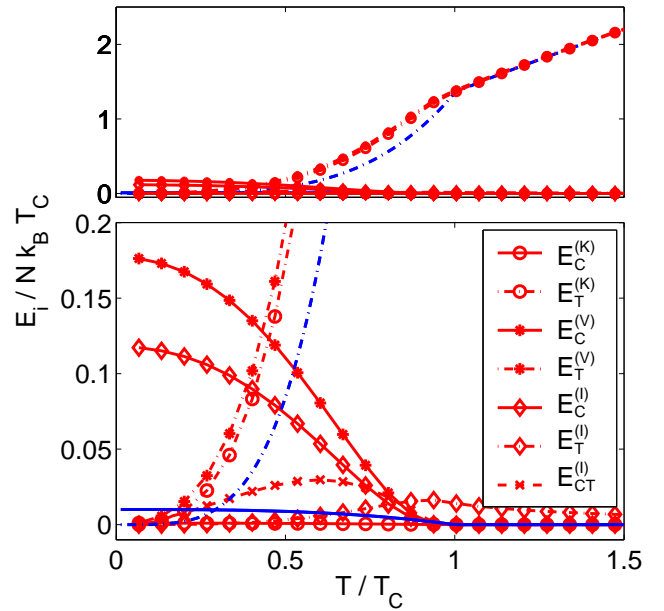}
\caption{(Color online) The temperature-dependent energy partitions of a Bose
gas in a spherically symmetric harmonic trap at $N=5\times 10^5$ and
$\omega=(2\pi)65$ Hz. The upper panel illustrates the overall feature while the
lower panel highlights the details of several small energy components.
In the lower panel, the two blue dotted lines refer to
the kinetic and (trap) potential energy for the thermal component
of an ideal Bose gas. They are essentially identical.
Similarly, the two solid blue lines denote the same energy components
as for the condensed part. The markers used for the various
energy components of an interacting Bose gas (in
red color) are defined in the legend.} \label{fig:fig3}
\end{center}
\end{figure}

The total energy of a Bose gas can expressed in terms of
its various partitions \cite{Giorgini97, Shi97}
\begin{eqnarray}
E &=& E^{(K)}_C + E^{(V)}_C + E^{(I)}_C\nonumber\\
&&+ E^{(K)}_T + E^{(V)}_T + E^{(I)}_T + E^{(I)}_{CT} \nonumber
\end{eqnarray}
with
\begin{eqnarray}
E^{(K)}_C &=&  \int d\vec r {\hbar^2\over 2M} |\nabla \Phi(\vec r)|^2, \nonumber
\\
E^{(V)}_C &=&  \int d\vec r\; V_{\rm ext}(\vec r) |\Phi(\vec r)|^2, \nonumber
\\
E^{(I)}_C &=& {g\over 2}\int d\vec r \; |\Phi(\vec r)|^4, \nonumber
\\
E^{(K)}_T &=& \int {d\vec rd\vec p \over (2\pi\hbar)^3} \; {p^2/(2M)\over
e^{\varepsilon(\vec p,\vec r)/k_BT}-1}, \nonumber
\\
E^{(V)}_T &=& \int d\vec r \; V_{\rm ext}(\vec r) n_T(\vec r), \nonumber
\\
E^{(I)}_T &=& g\int d\vec r \;  n_T^2(\vec r), \nonumber
\\
E^{(I)}_{CT} &=& 2g\int d\vec r \; |\Phi(\vec r)|^2 n_T(\vec r). \nonumber
\end{eqnarray}
$E^{(K)}_{C/T}$, $E^{(V)}_{C/T}$, and $E^{(I)}_{C/T}$ denote respectively the
kinetic energy, the trap potential energy, and the interaction energy
partitions of the condensed/thermal components. $E^{(I)}_{CT}$ denotes the
interaction energy between the condensate and the thermal component. Figure
\ref{fig:fig3} compares the temperature dependence for various energy partitions
of an ideal Bose gas (blue lines) with an interacting $^{87}$Rb gas (red lines)
in a spherical harmonic trap. For an ideal gas, only kinetic and trap potential
energy partitions are nonzero. In fact they are identically the same for either
the condensate or the thermal component. The total energy of the thermal
component exceeds that of the condensate when $T \gtrsim 0.3T_C$. For an
interacting gas, we see from Fig. \ref{fig:fig3} that the kinetic energy of the
condensate $E^{(K)}_C$ is negligibly small though it still decreases from a
nonzero value at $T=0$ to zero at $T=T_C$ if we inspect the curve carefully. At
low temperatures, $E^{(V)}_C/E^{(I)}_C \simeq 3/2$, which is a prediction from
the Thomas-Fermi approximation. At high temperatures, $E^{(V)}_C$ and
$E^{(I)}_C$ both approach zero as $T\rightarrow T_C$. Both kinetic and
potential energies $E^{(K)}_T$ and $E^{(V)}_T$ of the thermal cloud increase
rapidly as temperature increases ($T<T_C$), overwhelming the other energy
partitions near approximately $T_C/2$, and increase linearly for $T>T_C$.
$E^{(V)}_T$ is slightly larger than $E^{(K)}_T$ because of the repulsive atomic
interaction. They both approach their corresponding value for an ideal gas as
$T>T_C$, as shown in Fig. \ref{fig:fig3}. The interaction energy of the thermal
cloud $E^{(I)}_T$ reaches its maximum at $\sim\le T_C$ and decreases slightly
as $T$ increases. We also observe from Fig. \ref{fig:fig3} that the interaction
energy between the condensate and the thermal cloud $E^{(I)}_{CT}$ is largest
near $\sim 0.6T_C$ and decreases to zero at $\sim T_C$.

\begin{figure}
\begin{center}
\includegraphics[width=3.35in]{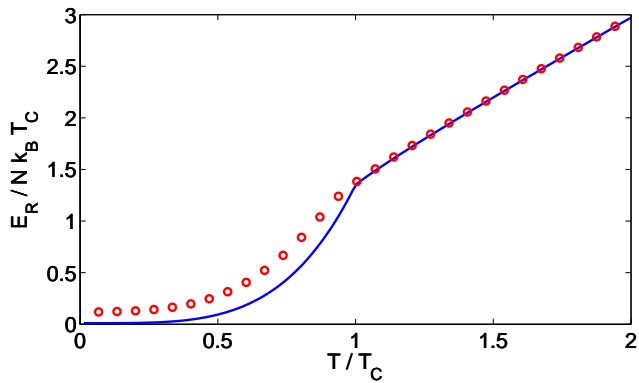}
\caption{(Color online) The temperature-dependent release energy of a
Bose gas for the same parameters as in Fig. \ref{fig:fig3}. The solid line is
for an ideal Bose gas while circles denote an
interacting $^{87}$Rb one. } \label{fig:fig4}
\end{center}
\end{figure}

Figure \ref{fig:fig4} illustrates the temperature dependence of the total release
energy of a Bose gas from a spherical harmonic trap, $E_{\rm R} =
E^{(K)}_C+E^{(I)}_C+E^{(K)}_T+E^{(I)}_T+E^{(I)}_{CT}$. We see that essentially
no difference exists between that of an ideal or an interacting gas. Comparing
Figs. \ref{fig:fig3} with \ref{fig:fig4}, we see the main feature of the release
energy at high temperatures again follow the characteristic temperature
dependence, i.e. $E_{\rm R} \propto (T/T_C)^4$ if $T<T_C$ and  $E_{\rm R}
\propto T/T_C$ if $T>T_C$, due mainly to the temperature dependence of the
kinetic energy of the thermal component $E^{(K)}_T$.

\section{Conclusion}
\label{sec:cl}

In conclusion, we have investigated theoretically the temperature dependence of
the average size of a trapped atomic Bose gas.
We found that the effective size can serve as a good
indication for the presence of BEC. Our mean field
theory gives several analytic results in the various temperature
limits, and allows for a careful comparison between the temperature dependence
of the various observable quantities between an ideal and
an interacting Bose gas. In addition, the simple temperature dependence of the
width of a trapped Bose gas can be potentially applied to
infer the temperature of an atomic gas cloud
\cite{Andrews96, Wang03}. A detailed analysis of the energy partitions show
that kinetic energy and trap potential energy of the thermal cloud represent
the main contribution to the energy of the system at high temperatures,
consequently they dominate the statistical properties of the system.
Furthermore,
our result provides an excellent theoretical support for the observation
of atomic BEC in Ref. \cite{Wang03}.\\

Part of this work was completed while one of us (L.Y.) was
a visitor at the Institute of Theoretical Physics of the Chinese Academy
of Sciences in Beijing, he acknowledges warm hospitality extended to him
by his friends at the Institute.
This work is supported by NSF, NASA, and the Chinese
Ministry of Education.

\end{document}